\documentclass[twocolumn,pra]{revtex4}

\usepackage{bm}
\usepackage{amsmath}
\usepackage{amsfonts}
\usepackage{graphicx}
\usepackage[caption=false]{subfig}
\usepackage{color}
\usepackage{amssymb}
\usepackage{hyperref}
\usepackage[ruled,linesnumbered]{algorithm2e}
\usepackage{multirow}
\usepackage{parskip}
\usepackage{tabularx}
\usepackage{psfrag,fancyhdr,layout,appendix}
\usepackage{verbatim}
\usepackage{framed}
\usepackage{amsthm}
\usepackage{booktabs}
\usepackage{placeins}
\usepackage{rotating}
\usepackage{changebar}
\usepackage{dcolumn}
\usepackage{multirow}
\usepackage{times}

\makeatother			

\newcommand{\ket}[1]{
    \ensuremath{\left|  #1 \right\rangle}\xspace}

\newcommand{\cnot}{\textsc{cnot}\xspace}

\newcommand{\PsiPlus}{\ensuremath{\ket{\Psi^+}}\xspace}

\newcommand{\myindent}{\hspace*{6pt}}

\newcommand{\mmlink}{\ensuremath{M\rightarrow M}\xspace}
\newcommand{\mmlinkleft}{\ensuremath{M\leftarrow M}\xspace}
\newcommand{\mimlink}{\ensuremath{M\rightarrow I \leftarrow M}\xspace}
\newcommand{\msmlink}{\ensuremath{M\leftarrow S \rightarrow M}\xspace}

\begin{document}

\title{Optimizing Timing of High-Success-Probability Quantum Repeaters}

\author{Rodney Van Meter, Takahiko Satoh, Shota Nagayama, Takaaki Matsuo and Shigeya Suzuki}
\affiliation{Keio University
  Shonan Fujisawa Campus, 5322 Endo, Fujisawa, Kanagawa, Japan}
\date{\today}

\begin{abstract}
  Optimizing a connection through a quantum repeater network requires
  careful attention to the photon propagation direction of the
  individual links, the arrangement of those links into a path, the
  error management mechanism chosen, and the application's pattern of
  consuming the Bell pairs generated.  We analyze combinations of
  these parameters, concentrating on one-way error correction schemes
  (1-EPP) and high success probability links (those averaging enough
  entanglement successes per round trip time interval to satisfy the
  error correction system).  We divide the buffering time (defined as
  minimizing the time during which qubits are stored without being
  usable) into the link-level and path-level waits.  With three basic
  link timing patterns, a path timing pattern with zero unnecessary
  path buffering exists for all $3^h$ combinations of $h$ hops, for
  Bell inequality violation experiments (B class) and Clifford group
  (C class) computations, but not for full teleportation (T class)
  computations.  On most paths, T class computations have a range of
  Pareto optimal timing patterns with a non-zero amount of path
  buffering.  They can have optimal zero path buffering only on a
  chain of links where the photonic quantum states propagate counter
  to the direction of teleportation.  Such a path reduces the time
  that a quantum state must be stored by a factor of two compared to
  Pareto optimal timing on some other possible paths.
\end{abstract}

\maketitle

\section{Introduction}

Distributed, entangled quantum states are useful for a variety of
purposes, including creation of provably secure, classical random bits
to be used as keys for encryption~\cite{ekert1991qcb}; distributed
physical reference frames for clock synchronization and optical
long-baseline interferometry for
astronomy~\cite{PhysRevLett.87.129802,chuang2000qclk,PhysRevLett.109.070503};
and distributed quantum
computation~\cite{crepeau:_secur_multi_party_qc,broadbent2009universal,Tani:2012:EQA:2141938.2141939}.

In the early days of quantum information research, Bennett \emph{et
  al.} recognized that the component qubits of an entangled state may
be held at different times in different locations, which they called
\emph{time-separated Bell pairs}~\cite{PhysRevA.54.3824}.  We can
apply this principle to group applications into three categories,
depending on their timing patterns: Bell inequality violations and QKD
(B class), which allow post-selection of successful trials; Clifford
group computations, such as creating cluster states (C class), which
allow post-operation application of Pauli frame corrections; and full
teleportation and non-Clifford computations (T class), which require
completion of all Pauli corrections before proceeding.  Each class
places different demands on the timing of operations at each node, and
ultimately affects the efficiency of a repeater network.  These three
patterns are summarized in Tab.~\ref{tab:apps}.

\begin{table*}
\begin{centering}
\begin{tabular}{| c | l || c | c | }
  \hline
Class & Name  & reception success notification & Pauli frame propagation \\ \hline
B & Bell (Bell inequality violations, QKD) & I & I \\ \hline
C & Clifford group computation & W & I \\ \hline
T & General (non-Clifford group) computation, teleportation & W & W \\ \hline
\end{tabular}
\end{centering}
\caption{Application usage patterns. The label 'W' indicates that the
  class must wait for the condition to be fulfilled before proceeding to
  the next step using the quantum memory, while 'I' indicates that the
  application may proceed immediately, without waiting.}
\label{tab:apps}
\end{table*}

Creating distributed quantum states over a distance is the work of a
chain of quantum repeaters~\cite{dur:PhysRevA.59.169}.  A
\emph{quantum repeater} has three functions: creating base-level
entanglement (typically a Bell pair) across a distance; coupling two
entangled states to lengthen entanglement; and managing errors.
Repeater chains may use purification as an error detection
mechanism~\cite{dur:PhysRevA.59.169}, or quantum error correction
(QEC)~\cite{knill96:concat-arxiv,PhysRevA.79.032325,PhysRevLett.104.180503,muralidharan15:_ultrafast-generations}.
In this paper we focus on QEC-based repeaters, which allow one-way
classical communication to be effective.  Links and nodes will be
organized into large-scale networks, which in turn may connect into a
quantum
Internet~\cite{kimble08:_quant_internet,van-meter14:_quantum_networking,PhysRevA.93.042338}.
A chain of repeaters selected from among the network's links for the
purposes of communicating between two given nodes is a \emph{path}.

The performance of a path of quantum repeaters, in terms of end-to-end
Bell pair generation rate and fidelity, is largely determined by the
efficient use of buffer memory in the repeater nodes~\footnote{Here we
  focus on repeaters that use some form of stationary memory at the
  nodes; this analysis does not apply to all-optical repeater
  proposals.}.  Efficiency in turn is driven by the success
probability per entanglement trial (determined by link hardware and
channel loss), trial repetition rate (determined by link architecture
and channel length), timing of the use of path elements (the focus of
this article), and the timing of consumption of the end-to-end Bell
pairs (determined by the application communication pattern, above).

The arrangement of components in a repeater link dictates the
propagation direction and travel distance of photons in the channel.
Some link timing patterns have a polarity, or direction of
propagation, while others are symmetric, with two photons originating
in the middle or meeting in the middle.

A path can be composed of links oriented in a large number of possible
patterns: $2^h$ for $h$ hops, or $3^h$ if we allow interspersed
unpolarized links.  If the probability of photon reception is low,
operation will be highly asynchronous, but when the probability is
high, we can coordinate the relative timing of operations on each link.
In this article, we present our analysis of several classes of path
timing patterns, focusing on paths with polarity to analyze systems
with the goal of minimizing memory buffering time and maximizing
performance.

We find that B and C class applications can always achieve the minimum
buffer timing of a single end-to-end round trip (or slightly less for
B class on many paths), including the time awaiting confirmation of
entanglement across each link.  For T class applications, an optimal
timing pattern exists only for a path composed of links with photons
propagating counter to the direction of teleportation.  In general,
the \emph{minimum} amount of time that the qubits building a
distributed quantum state dwell in memory, including awaiting the link
level acknowledgement, varies by up to a factor of two, depending on
path link pattern, error management mechanism, and application
pattern.

These results advance our understanding of how to engineer quantum
repeater networks for deployment in the real world, by establishing
the ``best best-case scenario'' and the ``worst best-case scenario''
for performance.

\section{Elementary Timing Patterns}

Before discussing the composition of link timing choices to create
path timing sequences, we must discuss in more detail the individual
link level timing patterns, the relationship to error management
(purification or correction), and the relationship to how the
application uses entanglement.

In all of the examples in this article, we will assume ``forward''
teleportation, drawn left to right in the diagrams.  Bell inequality
violation experiments and QKD do not have such end-to-end
application-level polarity.  In some cases, this allows them to behave
apparently atemporally, with qubits measured and released at the right
end of a connection before photons are created at the left end.  The
resulting classical data is not useful until Pauli frame corrections
are received~\cite{bertlmann12}, but our focus in this article is the
consumption of quantum resources (especially memory time).

\subsection{Link Level Timing}

\begin{figure}
\includegraphics[width=8cm]{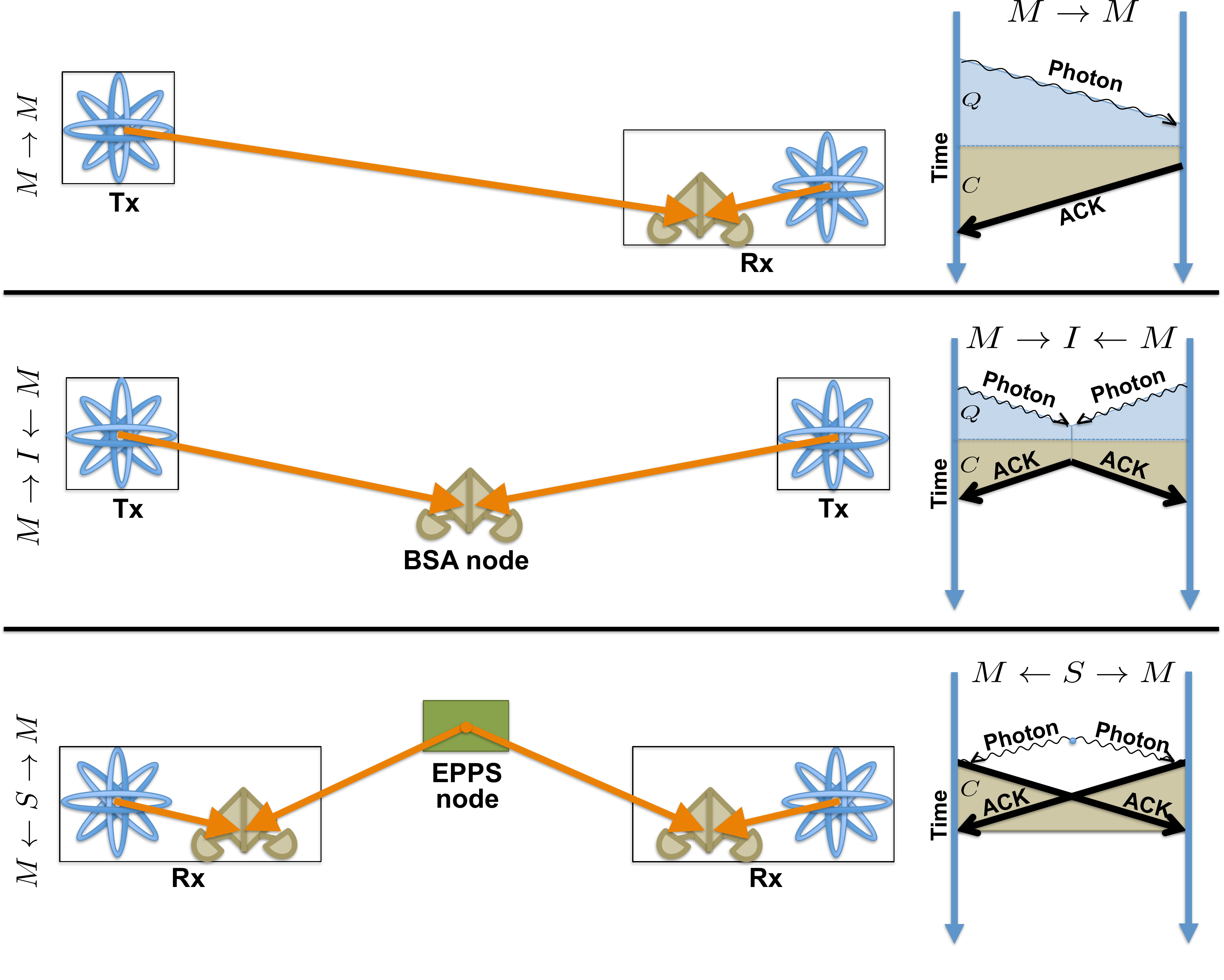}
\caption{Hardware arrangements for \mmlink, \mimlink, and \msmlink
  links.  Stationary memories are represented by the atom symbols.
  For the \mmlink link, the transmitting (left) end of the link should
  have more memories than the receiving (right) end, in proportion to
  the entanglement success probability.  Timing patterns and memory
  buffering times are shown in the diagrams on the right. The
  \mmlink link has transmitter-receiver polarity, evidenced
  by the trapezoid shape.  Shaded areas marked $Q$ indicate delay
  during quantum signal propagation, while shaded areas marked $C$
  await classical signal propagation.}
\label{fig:link-and-timing}
\end{figure}

There are three major link architectures classified by their timing
pattern, \mmlink (and \mmlinkleft), \mimlink, and \msmlink, summarized
in Tab.~\ref{tab:links} and illustrated in
Fig.~\ref{fig:link-and-timing}~\cite{van-meter14:_quantum_networking}.
We will use $t$ for one-way link propagation times, and $T_E$ for
one-way propagation end-to-end along an entire path.  When necessary
to discuss the latency of a single link within a path, we will refer
to $t_j$ for the $i$th link.

\mmlink (\mmlinkleft) can be considered to be a sender-receiver
(receiver-sender) arrangement.  A link may be composed of an emitter
at the transmitting end, and another emitter coupled with an
interference apparatus at the receiving end such that the combination
behaves as a
receiver~\cite{hong-PhysRevLett.59.2044,PhysRevA.59.1025,duan2001ldq,duan:RevModPhys.82.1209}.
Other physical mechanisms, such as qubus links, also exhibit the
\mmlink pattern~\cite{spiller05:_qubus}.

\mimlink uses two transmitters
and a Bell state analyzer (BSA, e.g., Hong-Ou-Mandel dip) positioned
halfway in between the two to erase which-path information, leaving
the two stationary qubits in an entangled
state~\cite{hong-PhysRevLett.59.2044,PhysRevA.59.1025,duan:RevModPhys.82.1209}.
This approach has seen substantial experimental success~\cite{hucul2014modular,sipahigil:PhysRevLett.108.143601,bernien2013heralded}.

\msmlink uses an entangled photon pair source (EPPS) at the midpoint
of the link, sending one photon toward each of two
receivers~\cite{jones2015design}.  Satellite-based repeater links are
also \msmlink, although the buffering latency is determined by the
terrestrial classical communication time rather than the actual
propagation time from the
satellite~\cite{aspelmeyer2003ldq,PhysRevA.91.052325,PhysRevLett.94.150501}.

Success probability per trial remains a critical Achilles heel in
experimental work, with values around
$10^{-5}$~\cite{hucul2014modular}.  The no-cloning
theorem~\cite{wootters:no-cloning,PhysRevLett.78.3217} dictates that
attempts to directly transmit quantum states encoded in an error
correcting code require that more than fifty percent of the qubits
arrive, $P_r > 0.5$, where $P_r$ is the probability of successful
reception of the photon.  Because technology sufficient to achieve
this level of success remains far off, we must instead first create
entangled Bell pairs, before attempting to use them for communication.
We refer to a link architecture that confirms the creation of Bell
pairs as \emph{acknowledged entanglement creation}, and the supporting
classical messaging protocol as
AEC~\cite{van-meter07:banded-repeater-ton,aparicio11:repeater-proto-design}.
We make the distinction from link-level heralding, as heralding is a
local event confirming successful entanglement, where that information
may be used locally or transmitted to the partner.

We must buffer the qubits at each end until the necessary classical
messages can arrive.  All three link architectures require decoherence
lifetime sufficient for retaining high fidelity after a full round
trip time ($2t$), but the location of that memory buffering and the
reason for it varies.  During the time period in which the memory
qubit is entangled with an in-flight photonic state, or we are
awaiting acknowledgement of entanglement success or failure, we refer
to the memory as being in the \emph{half-entangled}
state~\cite{aparicio11:repeater-proto-design}.  Tab.~\ref{tab:links}
lists the half-entangled quantum time (Q) during which a stationary
qubit is entangled with an in-flight photon and the half-entangled
classical time (C) during which the photon has arrived or failed to
arrive, but we are awaiting confirmation.  These times are marked in
Fig.~\ref{fig:link-and-timing}.  In \mmlink, all of the time is at
the transmitting end, whereas in \mimlink and \msmlink the buffering
time is evenly split between the two ends.  Assuming the decoherence
process is memoryless and the memory types at both ends are identical,
the net decoherence on the resulting Bell pair caused by storing a
single memory for time $2t$ is the same as for storing two memories
for time $t$ each.

The figure shows a single qubit at each node, but we can enhance
performance by using more qubits at each end.  The analysis here
assumes that the two arms of \mimlink and \msmlink links are balanced
in both latency and memory capacity, but that need not be so.

\begin{table*}
\begin{centering}
\begin{tabular}{| p{2.5cm} | l || c | c | c | c | c | c | c | l | }
  \hline
Link & long name  &  \multicolumn{3}{|c|}{``left'' end} &
\multicolumn{3}{|c|}{``right'' end} & & \\ \hline
&  &  Tx/Rx & Q & C & Tx/Rx & Q & C & L:R buffer
balance & rep rate \\ \hline\hline
\mmlink & SenderReceiver & Tx & $t$ & $t$ & Rx & 0 & 0 & $P_r^{-1}:1$ & 1/RTT \\ \hline
\mmlinkleft & ReceiverSender & Rx & 0 & 0 & Tx & $t$ & $t$ & 1:$P_r^{-1}$ & 1/RTT \\ \hline
\mimlink & MeetInTheMiddle & Tx & $t/2$ & $t/2$ & Tx & $t/2$ & $t/2$ &
1:1 & 2/RTT \\ \hline
\msmlink & MidpointSource & Rx   & $0$ & $t$ & Rx  & $0$ & $t$ & 1:1 & many/RTT
 \\ \hline
\end{tabular}
\end{centering}
\caption{Link architectures.  Each end (conventionally ``left'' and
  ``right'') is identified as transmitter (Tx) or receiver (Rx).  Q
  and C are quantum and classical signal propagation wait times, the
  half-entangled time in which a buffer memory 
  is awaiting confirmation of entanglement success with a partner.
  rep rate is the rate at which a given buffer qubit can be used in an
  entanglement attempt.  RTT is the link signal round trip time, $2t$.}
\label{tab:links}
\end{table*}

\begin{table*}
\begin{centering}
\begin{tabular}{| l | l | c | p{2.1in} | p{1.85in} |}
  \hline
  Regime & Reception Probability Range & D/G & Description &  Suitable timing architecture \\\hline
  low probability & $P_r \lessapprox \frac{2n}{N_{T}}$ & G & Prob. too low for
  coordinated operation & Fully async, ACKed 1-EPP or 2-EPP building E2E Bell pairs \\
  {\bf high probability} &
  $\frac{2n}{N_{T}} \lessapprox P_r < 0.5$ & G &
  Using many transmitters, high prob. of enough qubits being received
  to compose full QEC block size, but not above no-cloning loss limit
  & ACKed, optimized 1-EPP building E2E Bell pairs \\
  {\bf high prob. (extended)} &
  $0.5 < P_r < \alpha$ & G &
  Above no-cloning loss limit, but not yet practically so
  & ACKed, optimized 1-EPP building E2E Bell pairs \\
  very high probability &
  $\alpha < P_r < 1.0-\epsilon$ & D &
  Above no-cloning loss limit; more than half the qubits will arrive,
  allowing QEC reconstruction of state
  & fully 1-EPP, direct transmission of state in QEC block \\
  perfect & $1.0-\epsilon \le P_r $ & D &
  Error management does not need to account for loss &
  direct transmission (gate error rates permitting) \\\hline
\end{tabular}
\end{centering}
\caption{Matching probability regimes with appropriate entanglement
  purification patterns.  $P_r$ is the photon reception or
  entanglement creation probability.  $N_{T}$ is the number of
  transmitter qubits 
  in an \mmlink link, and $n$ is the block size in an [[$n$,$k$,$d$]] quantum
  error correction code.  $\alpha$ is chosen to assure low
  probability of $< \frac{\lfloor(n+1)/2\rfloor}{n}$ successes in
  every block operation across the entire path
  (see Sec.~\ref{sec:direct-prob}).  In
  the ``D/G'' column, ``D'' denotes direction transmission of valuable
  quantum data, while ``G'' denotes creation of generic quantum states
  (e.g., Bell pairs) which later are used to teleport quantum data.
  In this paper, we focus on the high probability and high probability
  (extended) regimes.}
\label{tab:link-epp}
\end{table*}

\subsection{Error Management Timing}

In 1996, Bennett \emph{et al.}~\cite{bennett1996pne,PhysRevA.54.3824},
Knill and Laflamme~\cite{knill96:concat-arxiv}, and Deutsch \emph{et
  al.}~\cite{PhysRevLett.77.2818} introduced means of managing errors
in distributed Bell states.  An entanglement purification protocol
(EPP) takes several distributed, mixed quantum states and reduces them
to a smaller number of higher-fidelity states.  Some of the states are
assigned to be \emph{sacrificial test tools}, and others are assigned
to be the output states.  Essentially, an EPP begins with a
proposition, such as, ``This Bell pair is in the state \PsiPlus,'' and
uses the test tools to test the correctness of that proposition.

Among the several mechanisms, the primary distinction is whether
communication is \emph{one way} (1-EPP) or \emph{two way} (2-EPP).  A
2-EPP has three significant advantages: it is robust against loss of
qubits, its resource requirements are low, and the procedures are
simple.  A 1-EPP also has advantages: it can use any known quantum
error correcting code, it integrates well with notions of
error-corrected distributed computation, and most importantly, reduces
the need to buffer states while waiting for classical communication,
allowing a quantum network to propagate signals end to end in a
hop-by-hop fashion similar to classical networks.

The simplest 2-EPP is a $2\rightarrow 1$ protocol, in which one Bell
pair is used to test the parity of a second.  A \cnot gate is performed
between the qubits of the two Bell pairs at each node, the target
qubit is measured, and the results are exchanged.  If the measurement
results agree, confidence that the initial proposition about the state
is correct is strengthened.  If the results disagree, even the
remaining Bell pair must be discarded.

1-EPP mechanisms can either retain the received state in an error
corrected form, or reduce the initial set of Bell pairs to unencoded
Bell pairs, much like the simplest 2-EPP.  It is known that 1-EPP is
not robust against loss of more than 50\% of the Bell pairs.

Our natural preference is for the 1-EPP, due to its low buffering
requirements.  However, the number of qubits per node and the required
number of Bell pair creation successes in a single burst of trials is
high (see Sec.~\ref{sec:prob}).  For a surface code of distance $d$,
we need $d$ successes in the burst to achieve the optimal timing
addressed here.  For a CSS code of $[[n,k,d]]$, we need a burst of $n$
entanglement successes.

\subsection{Application Timing and Buffer Time}

\begin{figure*}[h]
  \subfloat[B class]{
    \includegraphics[width=5cm]{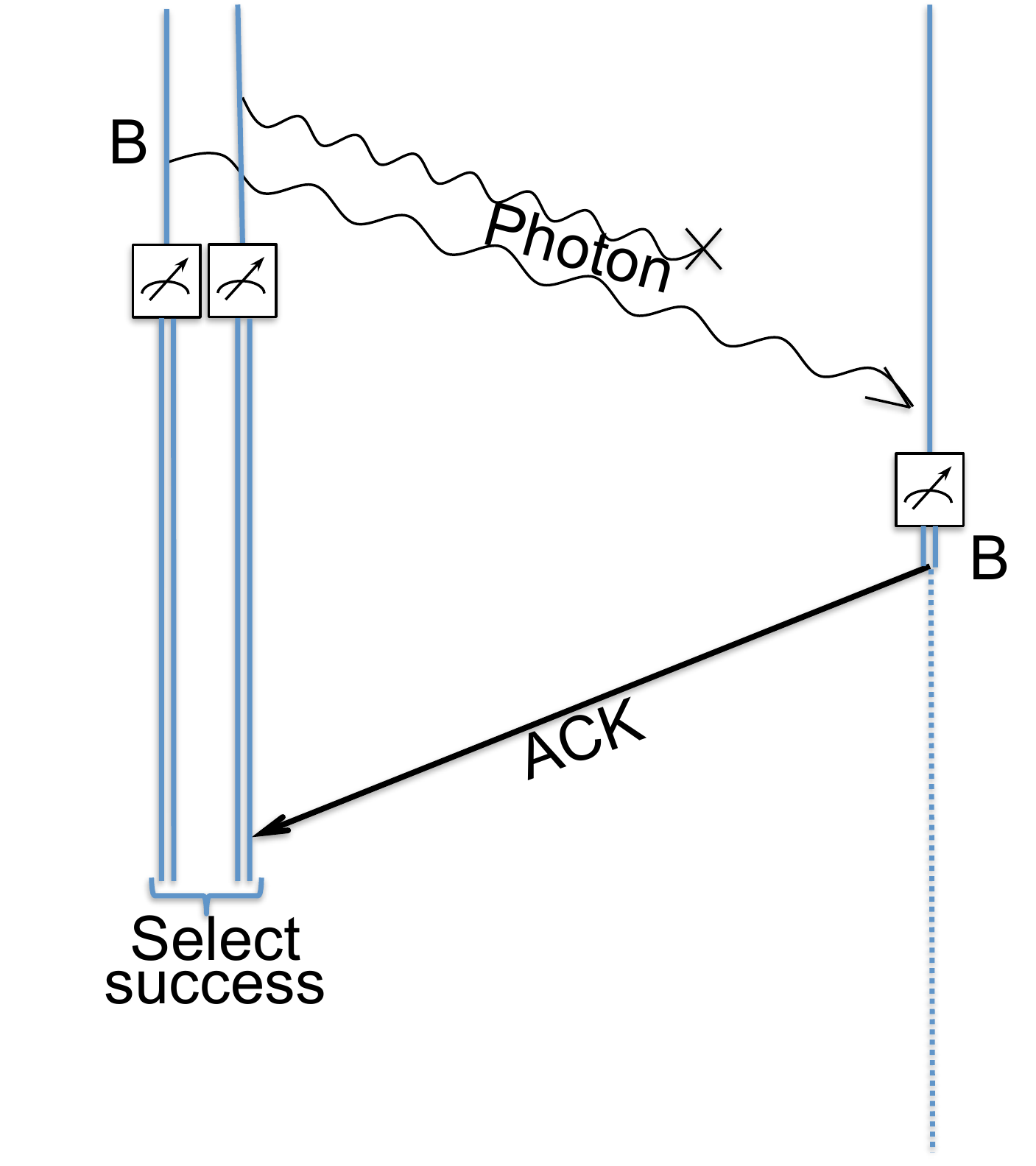}
    \label{fig:b-class}
    }
  \subfloat[C class]{
    \includegraphics[width=6cm]{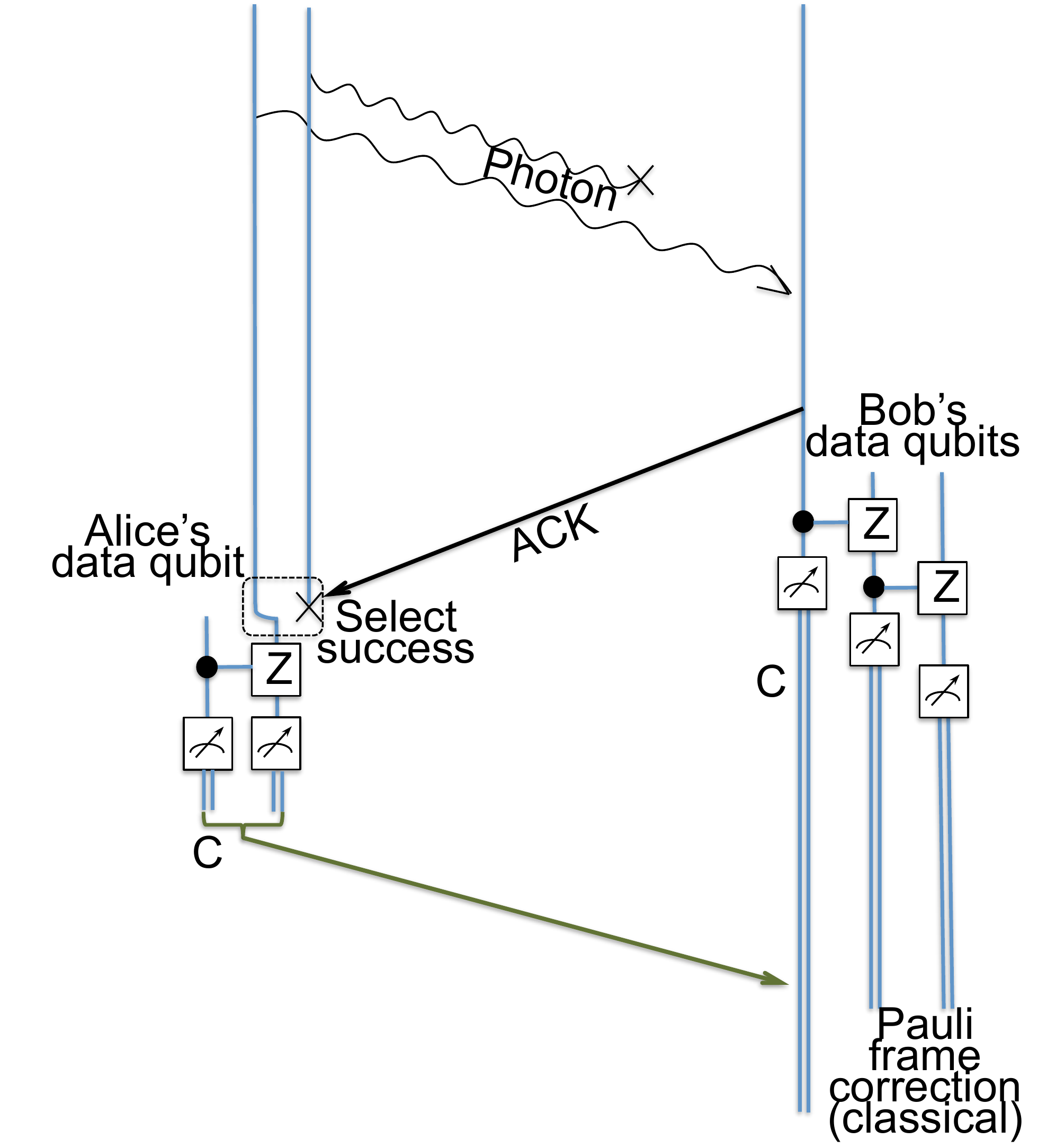}
    \label{fig:c-class}
    }
  \subfloat[T class]{
    \includegraphics[width=6cm]{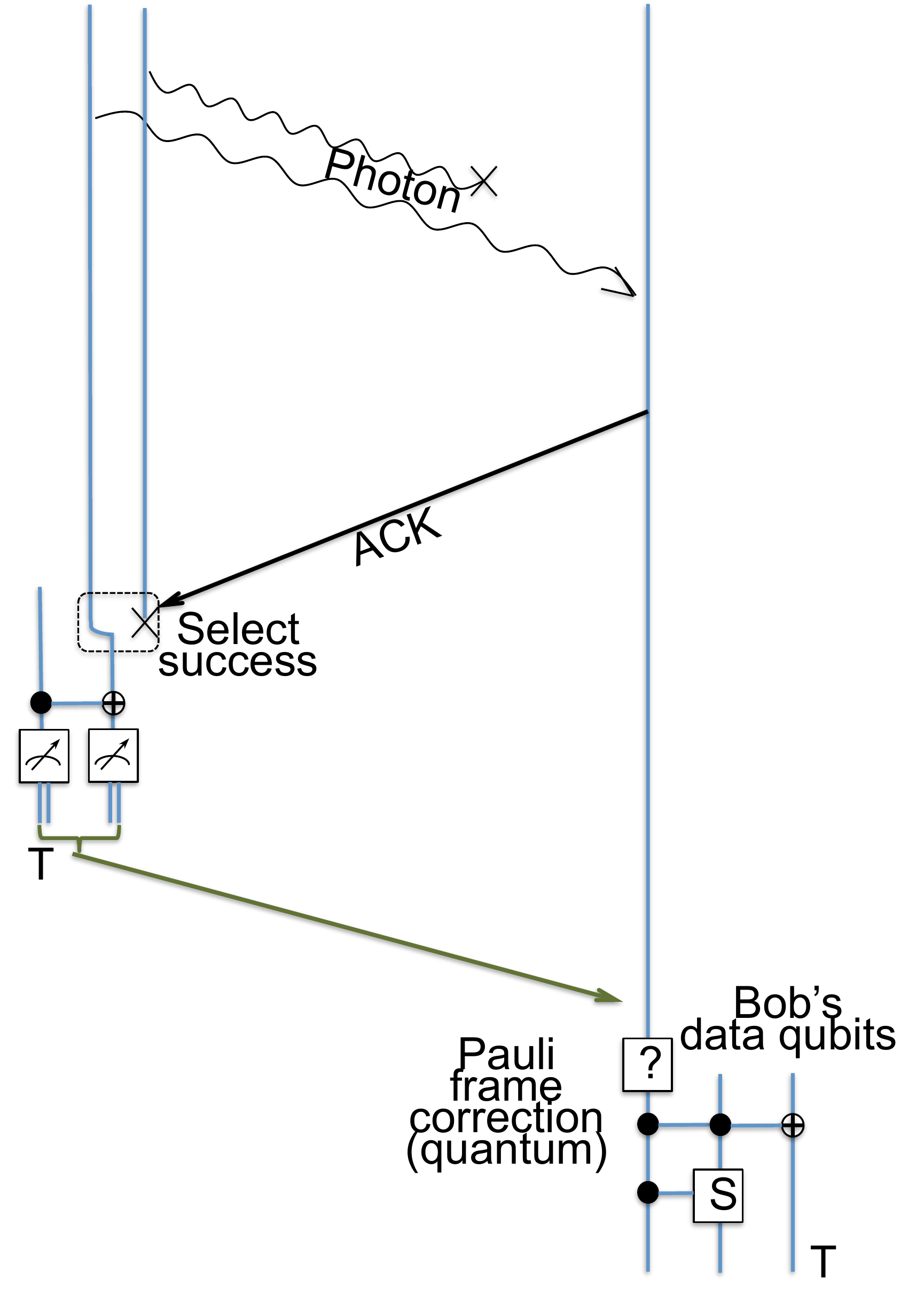}
    \label{fig:t-class}
    }
    \caption{Timing for application classes over a single
      forward-propagating link.}
\label{fig:app-classes}
\end{figure*}

Fig.~\ref{fig:app-classes} illustrates the timing and actions of the
three application categories listed in Tab.~\ref{tab:apps}.

Our waiting time can be divided into two categories: the link wait
time $T^H$ and the path wait time $T^P$.  The link wait time $T^H$ is
spent waiting for the classical entanglement success/failure
acknowledgment to arrive, also known as the \emph{half-entangled}
time.  For the C and T classes, which expect to execute additional
quantum operations on qubits at each end, we cannot use a qubit that
is entangled with a lost photon, as the photon loss leaves us in a
completely mixed state.  B class applications, in contrast, only
measure the qubit, and can post-select for entanglement success.  When
we have multiple qubits in flight, this acknowledgment signal serves
as a selection signal, telling us \emph{which} qubit (if any) to use,
as in the figure.  Over a single \mmlink hop, the half-entangled wait
times for the separate classes are
\begin{align}
T^H_B & = 0 \nonumber \\
T^H_C & = T^H_T = 2t.
\end{align}

During the path wait time, $T^P$, we cannot perform certain quantum
operations as we are awaiting either Pauli frame correction
information or the qubit-to-qubit matching information for
entanglement swapping (Sec.~\ref{sec:ent-swap}).  B class operations
can apply Pauli frame corrections to the classical measurement results
long afterwards, as with the photon reception post-selection.  C class
operations may proceed with Clifford group and measurement quantum
operations as soon as photon reception is confirmed, with Pauli frame
correction to follow later.  T class operations, which required
non-Clifford group operations, cannot proceed at the destination until
the Pauli frame correction is received and applied, giving
\begin{align}
T^P_B & = T^P_C = 0 \nonumber \\
T^P_T & = t.
\end{align}

In general, $T^H_C = T^H_T$ and $T^P_C = T^P_B$.  We define total
buffer times for B, C, and T classes to be $T_{B}$, $T_{C}$, and
$T_{T}$, respectively,
\begin{align}
T_B &= T^H_B + T^P_B = 0 \nonumber \\
T_C &= T^H_C +T^P_C = 2t \nonumber \\
T_T &= T^H_T + T^P_T = 3t,
\end{align}
representing the cumulative decoherence suffered by our quantum data.

\section{Path Timing Patterns}

Our goal is to determine the minimum achievable memory buffering time,
taking into account the application pattern and various link
arrangements.

\subsection{Building Long-Distance Entanglement}
\label{sec:ent-swap}

The original proposal for quantum repeaters used \emph{entanglement
  swapping}: Bob holds a Bell pair with Alice, and a second Bell pair
with Charlie, and performs a Bell state measurement using the two
qubits~\cite{PhysRevLett.71.4287,briegel98:_quant_repeater}.  This
disentangles Bob's two qubits and splices the original shorter Bell
pairs into one longer one.  However, in order to complete the
operation, a Pauli frame correction must be applied at one end, and
both ends must be informed of the completion of the operation.  The
propagation of these messages is one of the key constraints on system
performance.

The original proposal used entanglement swapping at the physical
level.  Jiang \emph{et al.} proposed using it at the logical
level~\cite{PhysRevA.79.032325}.  Fowler \emph{et al.} used the
surface code-specific approach of coupling together small segments of
the surface code to form a single, long, narrow
surface~\cite{PhysRevLett.104.180503}.

\subsection{Path Basics}

A \emph{path} is a concatenated sequence of links through a network
chosen to connect a source and
destination~\cite{van-meter:qDijkstra}.  The end-to-end latency of the
path is just the sum of the individual link latencies,
\begin{equation}
T_E = \sum_j t_j.
\end{equation}
This value has some relationship to the path performance and to the
decoherence suffered by stationary memories.  Exploring that
relationship, which is not straightforward, is the goal of this
paper.

If memory decoherence is a memoryless process, the sum of the buffer
times all along the path determines total decoherence.

In planning to eliminate unnecessary buffering, we care about two
factors:
\begin{enumerate}
\item the \emph{total amount} of unnecessary buffering, which
  determines the unnecessary decoherence suffered; and
\item the \emph{spatial distribution} of the unnecessary buffering,
  which determines the impact on aggregate network performance.
\end{enumerate}

To minimize buffer wait times, we coordinate action across the entire
path using one or more \emph{trigger} signals.  The trigger may
propagate left to right (forward) or right to left (reverse).  We may
also trigger from the middle of the path (inside out), or
simultaneously from both ends toward the middle (outside in).  Each
link may transmit photons in the same direction as the trigger
(co-propagating) or in the opposite direction (counter-propagating).
The design decision of the trigger and link propagation pattern has a
significant effect on buffer memory times.

Entanglement swapping information becomes available after the
neighboring link has confirmation of photon reception, either at the
sending or receiving end.  In Figs.~\ref{fig:figs} and
\ref{fig:hetero-patterns}, the path wait times $T^P$ (whether we are
waiting for Pauli frame information or entanglement swapping
information) are illustrated with red arrows outside of the trapezoids
representing link timing.  The link wait times $T^H$ are not directly
represented in the figures, but are understood to be along the long
edge of each trapezoid.

\subsection{Forward Propagation, Flat Timing}

\begin{figure*}[h]
  \subfloat[Forward propagation links, flat timing]{
    \includegraphics[width=8cm]{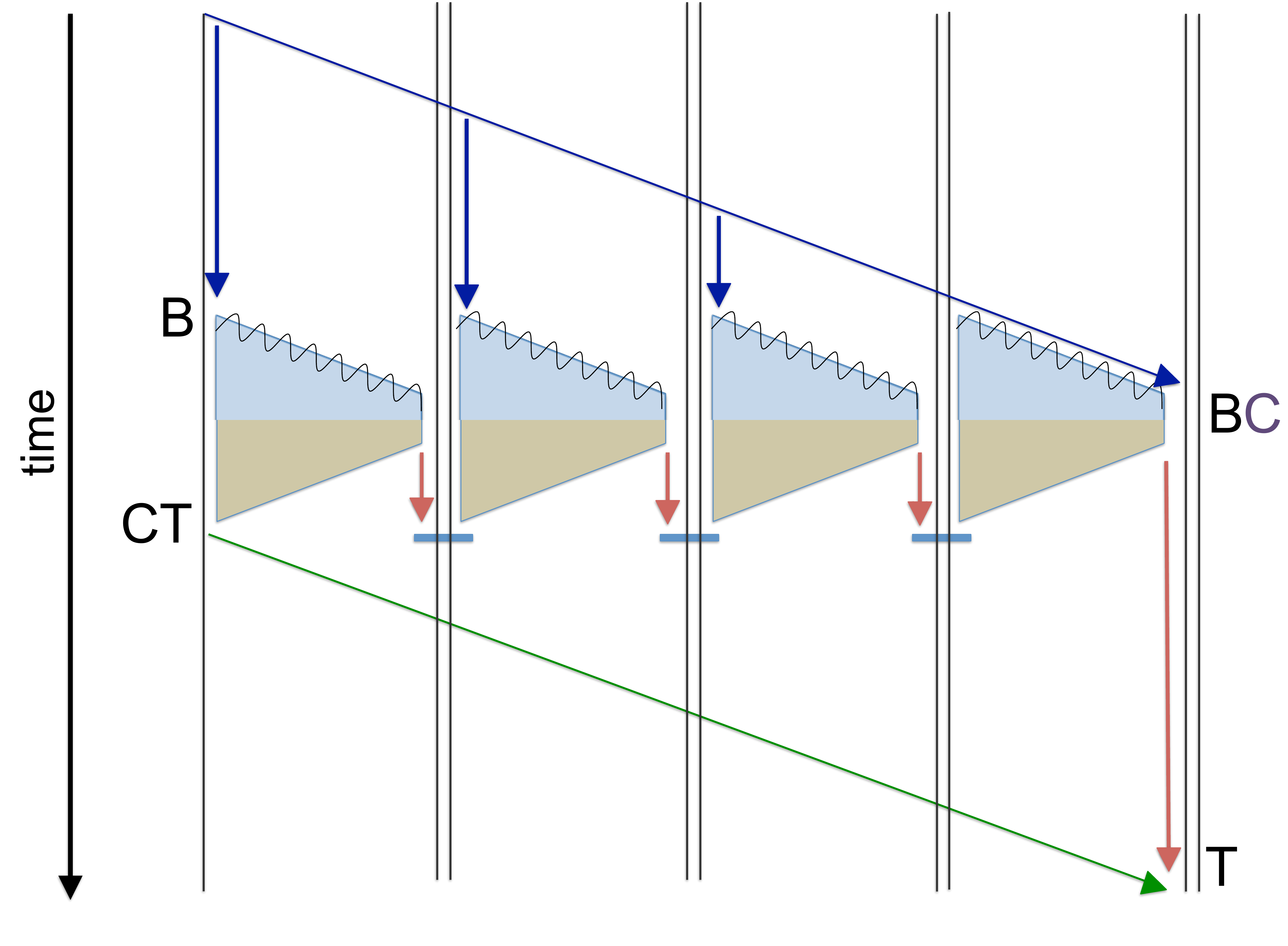}
    \label{fig:flat-fwd}
    }
  \subfloat[Forward triggered with varied-length hops, not Pareto optimal]{
    \includegraphics[width=8cm]{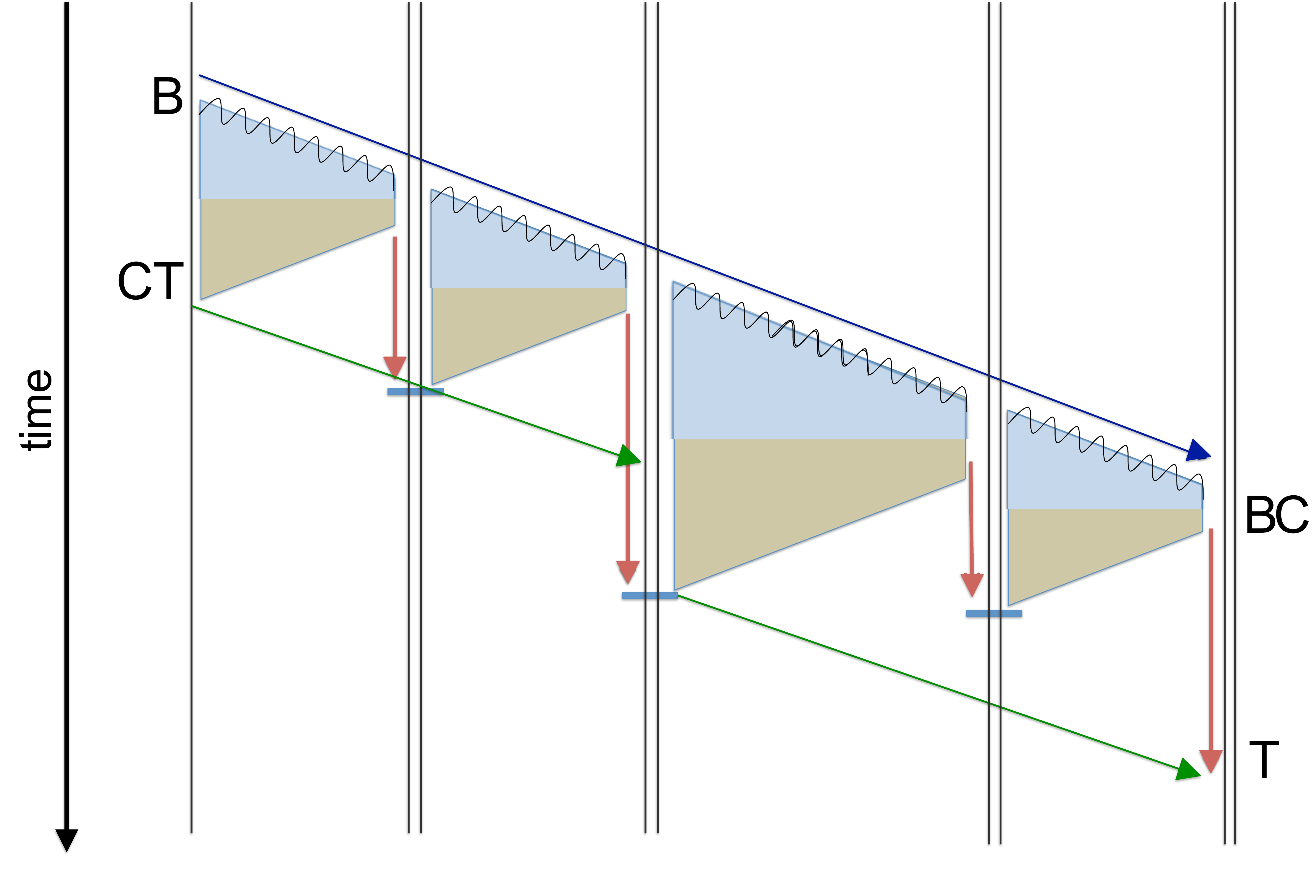}
    \label{fig:fwd-nonpareto}
    }\\
  \subfloat[Varied-length hops with corrected trigger]{
    \includegraphics[width=8cm]{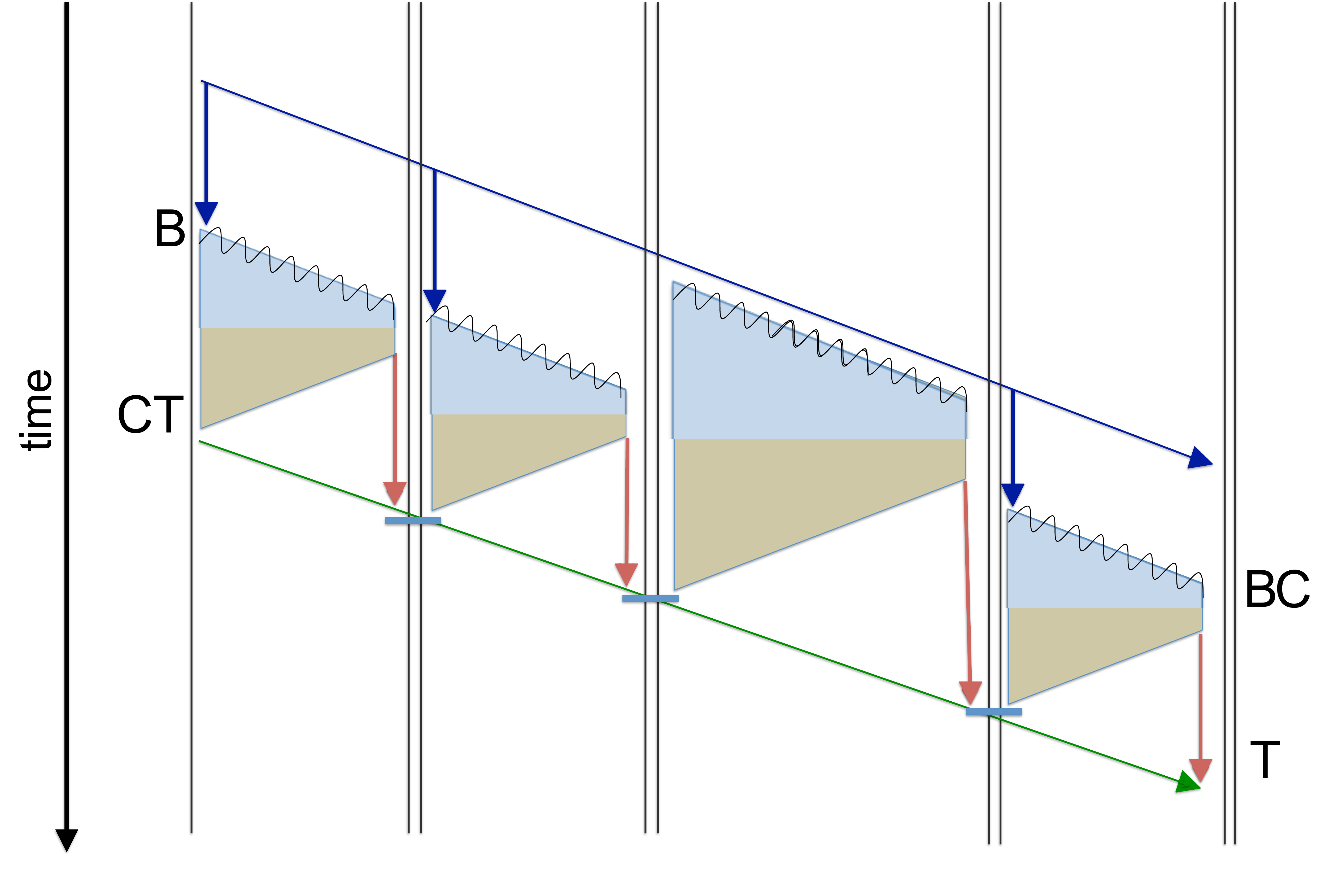}
    \label{fig:fwd-pareto}
    }
  \subfloat[Bell matched timing]{
    \includegraphics[width=8cm]{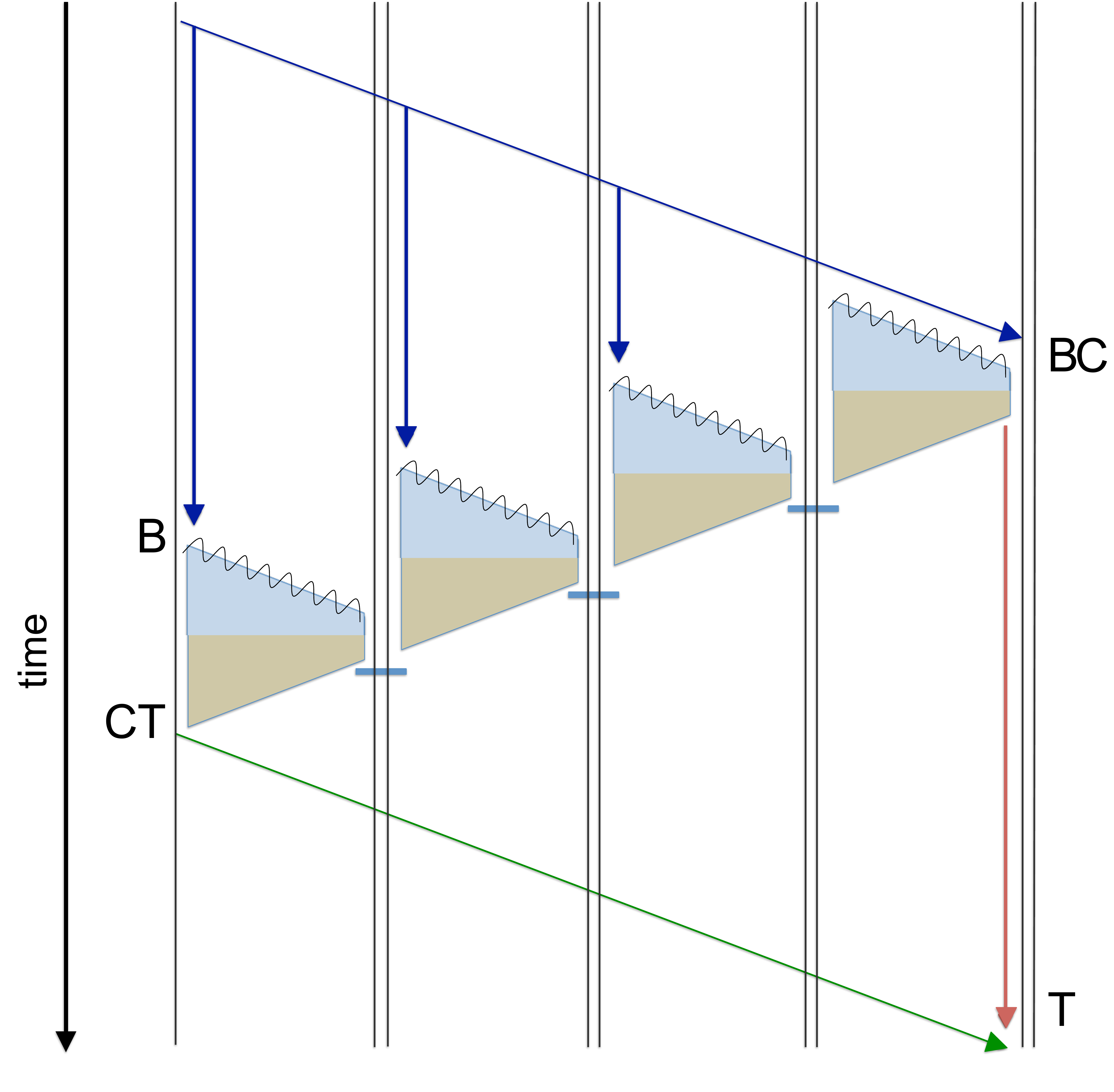}
    \label{fig:fwd-preferred}
    }
    \caption{Timing for forward operation with acknowledged
      link-level entanglement on a forward-propagating path.  Blue
      arrows are the timing trigger.  Green arrows are propagation of
      Bell state measurement results used as Pauli frame corrections.
      Red arrows indicate memory hold time waiting for BSA or final
      Pauli frame correction. (a) Flat-timed 1-EPP repeaters.  The red
      arrows here sum to one end-to-end round-trip time for the C and
      T classes. (b) Forward propagation with uneven link latencies
      and a simple trigger, causing the ``long pole'' to extend below
      the green arrow, violating the first Pareto optimality
      condition.  (c) The same path with corrected trigger, giving
      Pareto optimal timing.  (d) The Bell matched timing, optimal for
      all three classes.}
\label{fig:figs}
\end{figure*}

After selecting an error management mechanism, we can compose links
into a path, then add a path trigger pattern and application Bell pair
consumption information to project behavior.  This can be illustrated
graphically, and we can read the required buffer timings directly off
of each diagram, as in Figs.~\ref{fig:figs} and
\ref{fig:hetero-patterns}.

As an example, consider the basic operation of 1-EPP repeaters, as
described by Jiang \emph{et al.} and Fowler \emph{et al.}  The
analysis in both papers originally assumed synchronous operation
across all links, as illustrated in Fig.~\ref{fig:flat-fwd}.  While
this is not strictly necessary, it simplifies the exposition and basic
performance analysis.  The exact timing of the BSA at each node can in
fact be independent, as long as the BSA operations along the chain can
be mapped to the correct end-to-end session.

However, in the figure, the red arrows indicate time where only one
side of a planned entanglement swap is ready.  For this path, all
three application classes have unnecessary path buffering at every
node except the end points.  For the B and C application classes, this
is the only unnecessary buffering incurred, giving us
\begin{align}
T^P_B = T^P_C = \sum_{j=2}^{h}t_j = T_E - t_1
\end{align}
noting that the first hop is left out.  The T application
class requires us to wait at the right end until the Pauli frame
information arrives from the left end, giving
\begin{align}
T^P_T & = \sum_{j=2}^{h}t_j+t_h+T_E \nonumber\\
& = 2T_E, \text{when all links have the same $t_j$.}
\end{align}
or one end-to-end round-trip time.  B and C class buffering is evenly
spread along the path except for the last node.  The additional
buffering time for the T class is entirely at the right end.

\subsection{Forward Propagation, Other Timing Patterns}

The flat timing pattern is only one possible pattern for this fixed
arrangement of hops.  Different timing patterns are achieved by
sliding the link timing trapezoids up and down by adjusting the
trigger pattern.  Some moves result in a global reduction the
buffering time, others a global increase.  Some moves do not change
the global sum, but alter the location of the buffering.

Fig.~\ref{fig:fwd-nonpareto} shows an obvious but poor choice of
timing pattern, using a single forward trigger and immediately firing
each link on the trigger.  This pattern is worse than flat timing,
with twice the buffering for B and C classes and the same or worse
buffering for T class,
\begin{align}
T^P_B = T^P_C & = \sum_{j=2}^{h}2t_j \\
T^P_T & = \sum_{j=2}^{h}2t_j+ 2\operatorname{max}(t_j).
\end{align}

With unequal link latencies, the wait time at the middle node where
the two green arrows (Pauli frame corrections) do not line up
represents wasted time.  Sliding the two left links downward so that
the Pauli frame corrections form a single line will lower the global
buffer time in a somewhat ad hoc fashion, as shown in
Fig.~\ref{fig:fwd-pareto}.

Instead, to determine the actual minimum global buffer time and
establish a preferred timing pattern, we adjust the trigger timing on
each link to arrange for there to be no waiting on either side for
each Bell state measurement operation.  This will eliminate the red
arrows at every node except the right hand one, as in
Fig.~\ref{fig:fwd-preferred}.  In this case, we achieve
\begin{align}
T^P_B = T^P_C & = 0 \\
T^P_T & = \sum_{j=1}^{h}2t_j = 2T_E.
\end{align}

For B and C classes, this minimum value of zero on the forward
propagating path is achieved \emph{only} for this timing pattern.  For
the T class on this path, we have a continuum of Pareto optimal
timings, all with the minimal global buffering but with different
distributions of the buffering along the path.  Other Pareto optimal
solutions can be reached by moving any link or set of links up and
down such that (a) all Bell measurement operations stay above the
Pauli frame line (green arrow) propagating from the left edge, and (b)
every subset of trapezoids is in balance such that moving it up or
down as a group lengthens one red arrow the same amount as it shortens
another.  Noting that the red arrow is always on the side waiting,
condition (b) can equivalently be stated as no contiguous subsequence
of links starts with a red arrow on the left side of the leftmost link
and ends with a red arrow on the right on the right side of the
rightmost link.  Matching the timing of Bell measurements at each
junction is always optimal for B and C classes and always represents
one point (perhaps the only point) on the Pareto optimal frontier for
T class.

\subsection{Other Paths and Their Optimization}

In this section, we demonstrate a method for establishing an optimal
timing pattern for any given path on a fixed network, and for choosing
a link arrangement during the network design phase.

As noted in the introduction, each link can be left-to-right,
right-to-left, or unpolarized in transmission direction.  The link
timing options for a path composed of $h$ hops thus can exhibit $3^h$
different orientations.

In order to introduce the link polarity into
the latency, we refer to $\tau_j$ for the $j$th link, using
\begin{eqnarray}
\tau_j =  \begin{cases}
    0, & \text{if $\mimlink$ or $\msmlink$}.\\
    t_j, & \text{if $\mmlink$}.\\
    -t_j, & \text{if $\mmlinkleft$}.
  \end{cases}
\end{eqnarray}
For timing arranged with simultaneous arrival and immediate Bell state
measurement at each intermediate node, we obtain the following
equation for a path:
\begin{eqnarray}
0 \leq T^P_T = T_E + \sum_j \tau_j \leq 2T_{P}.
\label{eq:polarity-sum}
\end{eqnarray}

When we synchronize photon arrival timing on each link appropriately,
$T^P_B$ and $T^P_C$ always can be suppressed to the theoretical
minimum of zero.

\begin{figure*}[h]
\begin{center}
    \subfloat[Butterfly links, ridge fold timing]{
      \includegraphics[width=8cm]{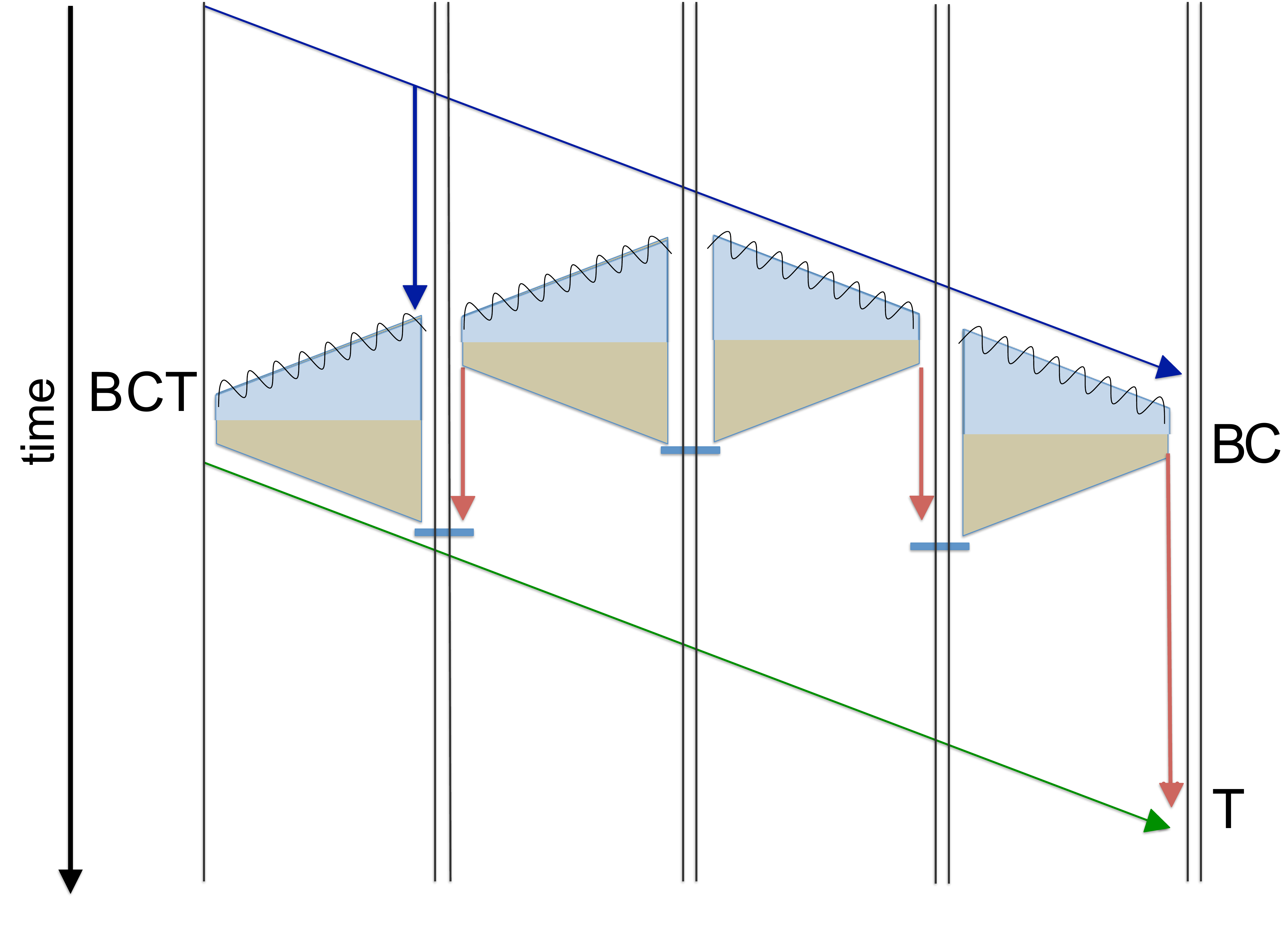}
      \label{fig:quasi-bfly}
    }
    \subfloat[Butterfly links, valley fold timing, Bell matched]{
      \includegraphics[width=8cm]{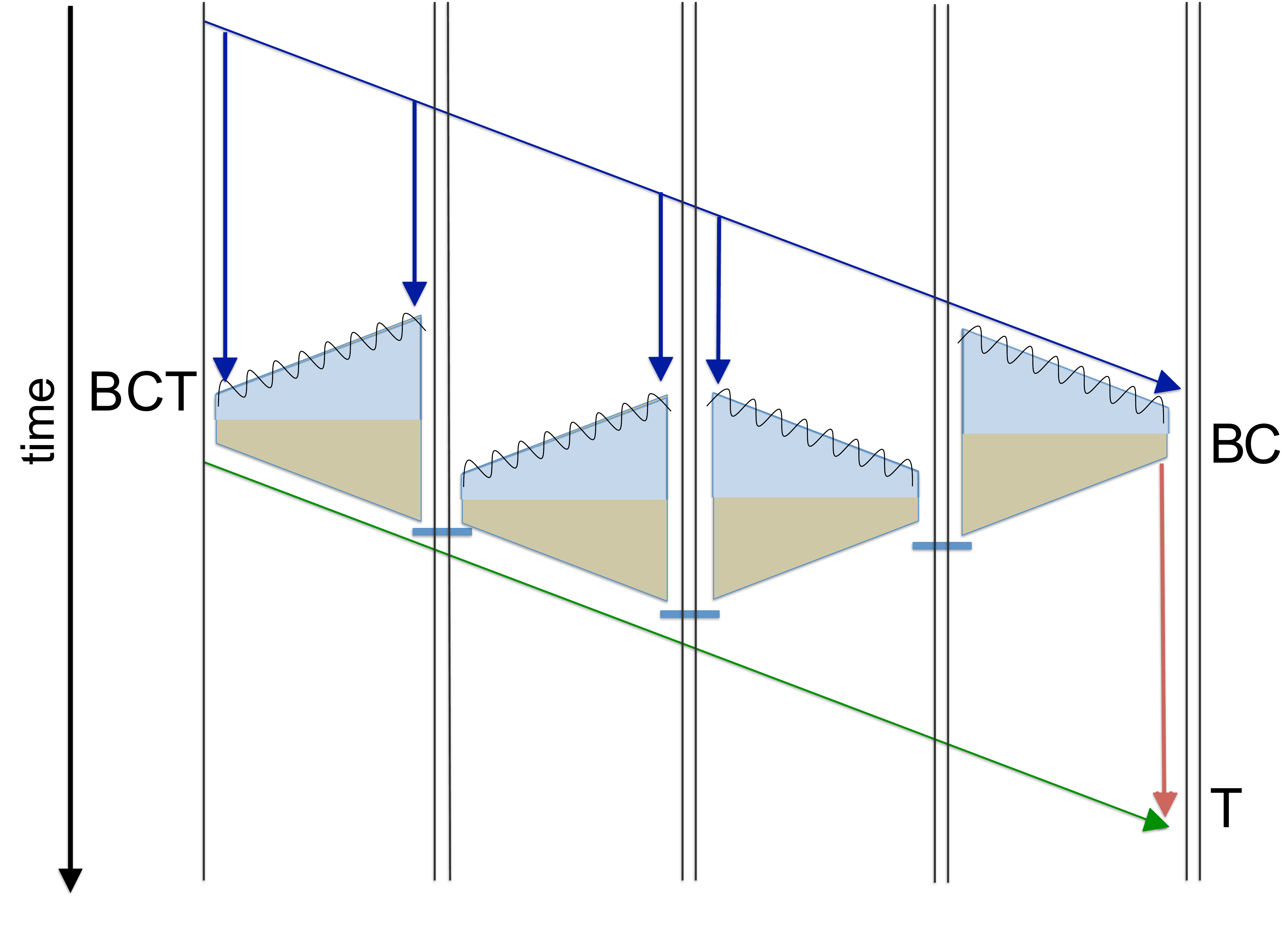}
      \label{fig:valley}
    }\\
    \subfloat[Inverted butterfly, Bell matched]{
      \includegraphics[width=8cm]{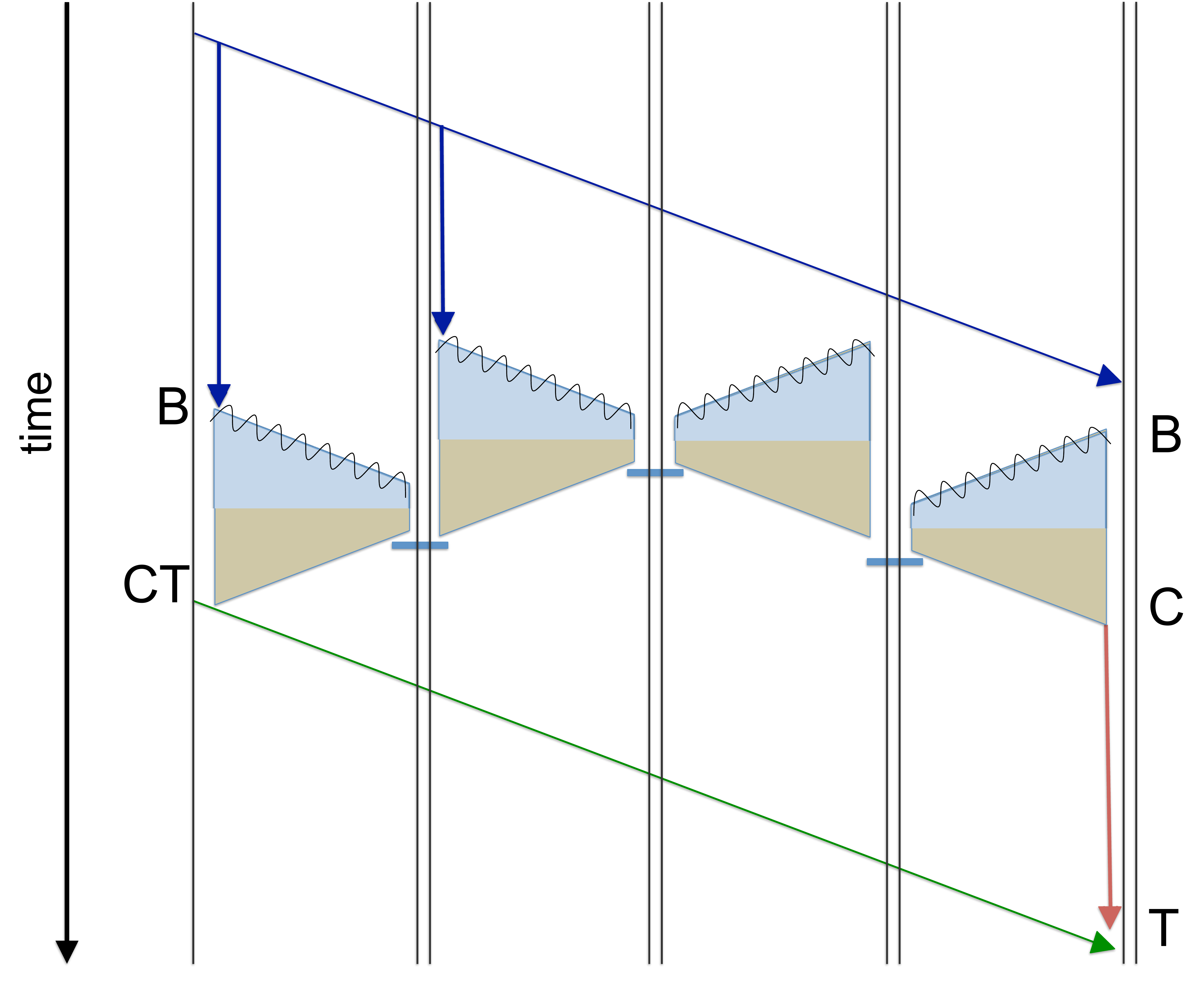}
      \label{fig:inv-bfly}
    }
    \subfloat[Reverse propagation, forward triggered]{
      \includegraphics[width=8cm]{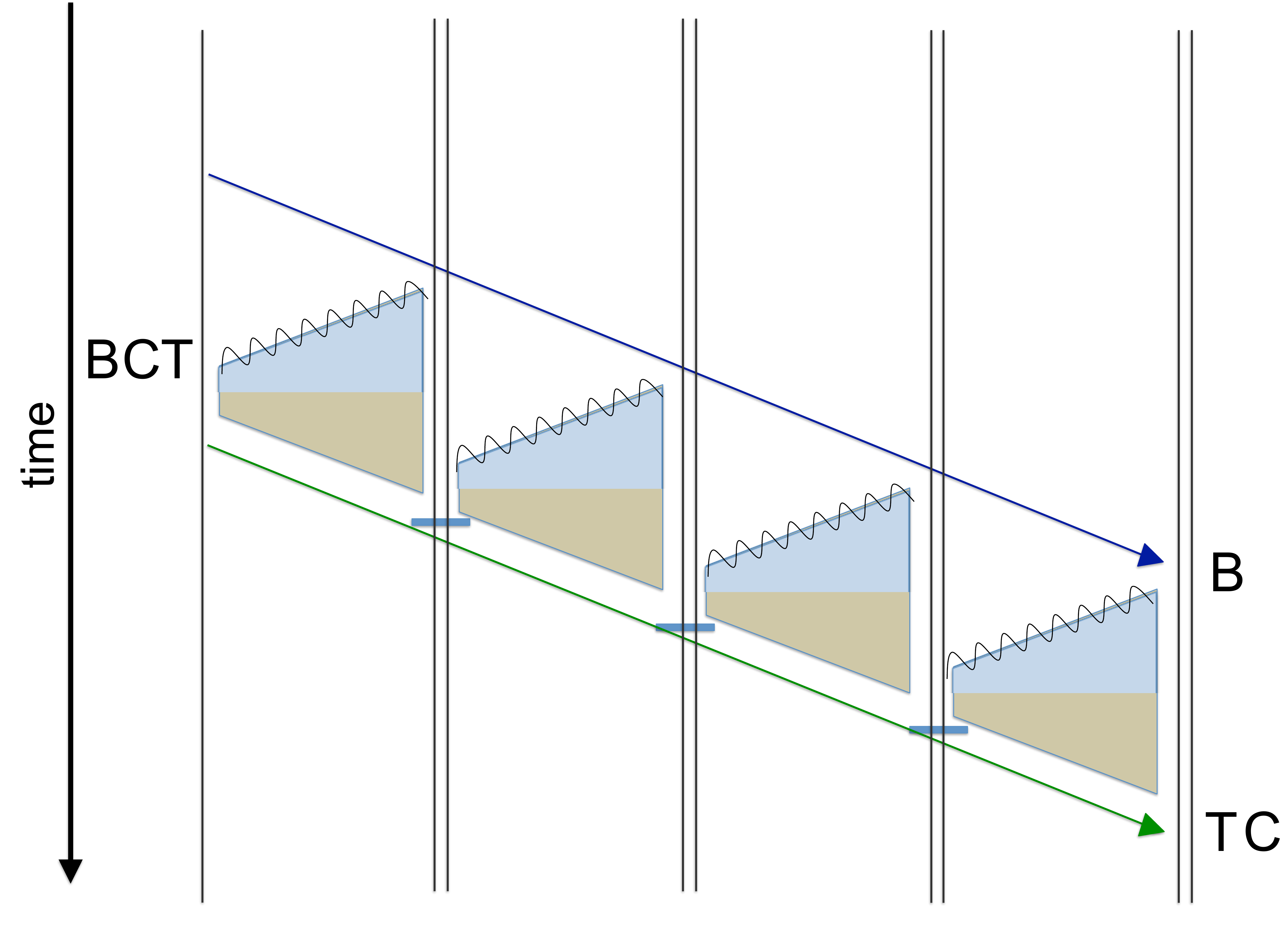}
      \label{fig:quasi-link-rev}
    }\\
    \subfloat[Mixed direction path, optimized timing]{
    \includegraphics[width=8cm]{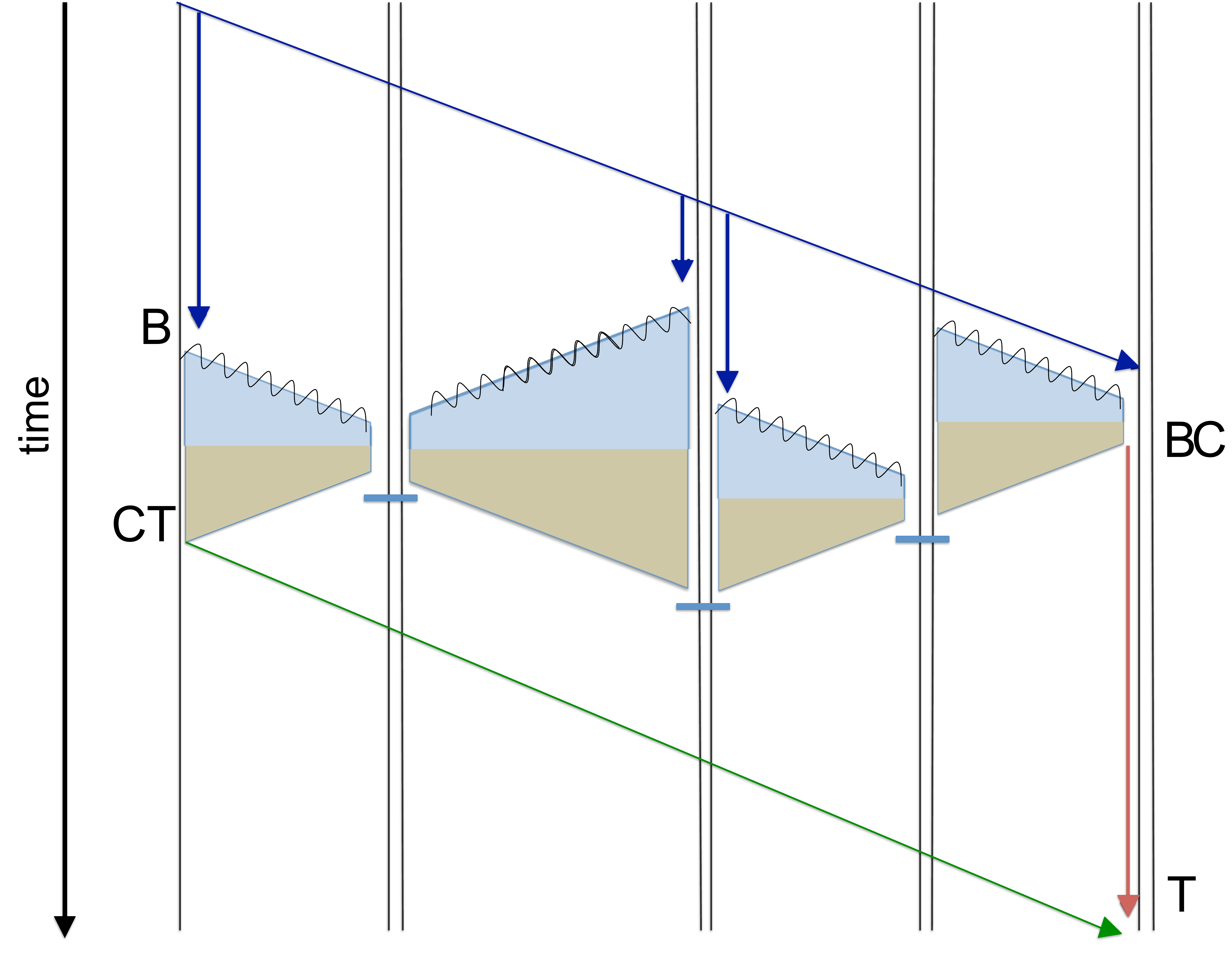}
    \label{fig:het1}
    }
\end{center}
\caption{Other link and timing patterns. (a) ``Ridge fold'' timing on
  butterfly link arrangement, with operation initiated from the
  midpoint of the path.  The second Pareto condition is violated
  because the middle two hops start and end with two matching red
  arrows.  (b) The same hardware configuration with ``valley fold''
  timing created by matching Bell measurements, achieving optimality
  for B class and Pareto optimality for C and T classes. (c) Inverted
  (inside out) butterfly, with Bell matched timing.  (d) Timing
  for counter-link propagation, with all of the links transmitting
  right to left to teleportation propagating left to right.  This is
  the only path achieving zero path buffering for all three
  classes.  (e) Mixed direction path using the matched Bell timing
  pattern, optimal for all three classes.}
\label{fig:hetero-patterns}
\end{figure*}

The ``butterfly'' arrangement of links with ``ridge fold'' timing,
shown in Fig.~\ref{fig:quasi-bfly}, has been
proposed~\cite{munro2010quantum}, but turns out not be Pareto optimal,
\begin{align}
T^P_T = 2T_E-2t_{\lfloor h/2\rfloor}-2t_{\lfloor h/2\rfloor+1}+\sum_{j=1}^ht_j \approx 3T_E.
\end{align}

Sliding the two middle trapezoids downward, we find the ``valley
fold'' arrangement (Fig.~\ref{fig:valley}), which is Pareto optimal
for B, C and T.  On this set of links, valley fold is the only optimal
timing pattern for classes B and C.  Within the rules established in
the prior section, the timing of the links in the right half of the
path can be adjusted while retaining Pareto optimality for class T,
adjusting the location of buffering in use,
\begin{align}
T^P_T = \sum_{j=\lfloor h/2\rfloor}^h2t_j \approx T_E.
\end{align}

The arrangement of Fig.~\ref{fig:inv-bfly} with the polarity of the
links inverted is superficially similar, but in fact results in rather
different buffering characteristics,
\begin{align}
T^P_B = T^P_C & = 0 \\
T^P_T & = 2T_E.
\end{align}

Fig.~\ref{fig:quasi-link-rev} shows a path where the photon
propagation is counter to the desired teleportation direction.  In
Eq.~\ref{eq:polarity-sum}, $\tau_j = -t_j$, giving $T^P_T = 0$.  This
is the only arrangement of links for which $T^P_B$, $T^P_C$, and $T^P_T$ are
all zero.  Thus, when designing a network, \emph{if the traffic
  pattern is known to be left to right along a fixed path}, this
is the preferred arrangement of links.

Fig.~\ref{fig:het1} shows an example of a pathway of mixed
polarization with Bell state measurement matched timing.  In this
fashion, all of the buffering is at the right end, class B and C
retain zero path buffer time, and class T is Pareto optimal.  It is
always possible to find such a timing arrangement.  Such irregular
paths will likely be the norm in operational networks.

\begin{table*}
\begin{centering}
\begin{tabular}{| l | c || c | c | c | c || c | c || p{2.7cm} |}
  \hline
  Path prop/ &&&&&&&& \\
  timing pattern & Matched? & $T^P_B, T^P_C$ & B,Copt? & $T^P_T$
  &Topt? & $T^H_B$ & $T^H_C, T^H_T$ & Buffering \\ \hline
  {\bf Forward} &&&&&&&& \\
  \quad flat & N & $T_E-t_1$ & N & $2T_E$ & P & $2T_E-2t_1$ & $2T_E$ & distributed, heavy
  at right end point \\
  \quad forward & N & $2T_E-2t_h$ & N & $2T_E$ & P & $2T_E-2t_1$ & $2T_E$ & evenly distributed \\
  \quad uphill & Y & 0 & Y & $2T_E$ & P & $2T_E-2t_1$ & $2T_E$ & all at right end point \\ \hline
  {\bf Butterfly} &&&&&&&& \\
  \quad ridge fold & N & $2T_E-2t_{\lfloor h/2\rfloor}-2t_{\lfloor h/2\rfloor+1}$ & N & $\approx 3T_E$ & N &  $2T_E$ &  $2T_E$ & distributed, heavy
  at right end point \\
  \quad valley fold & Y & 0 & Y & $T_E$ & P &  $2T_E$ &  $2T_E$ & all at right end
  point \\ \hline
  {\bf Inverted B'fly}  &&&&&&&& \\
  \quad ridge fold & Y & 0 & Y & $T_E$ & P & $2T_E-2t_1-2t_h$ &  $2T_E$ & all at right
  end point \\ \hline
  {\bf Reverse}  &&&&&&&& \\
  \quad forward & Y & 0 & Y & 0 & Y & $2T_E-2t_h$ &  $2T_E$ & none \\ \hline
  {\bf Mixed} &&&&&&&& \\
  \quad Bell matched & Y & 0 & Y & $T_{P} + \sum_j \tau_j$ & P & $\le 2T_E$ &  $2T_E$ & all at right end point \\ \hline
\end{tabular}
\end{centering}
\caption{Path timing patterns.  ``Matched?'' indicates whether the
  Bell measurement timings are matched.  ``opt?'' indicates whether
  the timing pattern is Pauli optimal for that path and class of
  applications. ``P'' indicates that the pattern is one 
  point on a Pareto optimal frontier with more than one point.
  ``Buffering'' describes the location of Pauli waits, if any.}
\label{tab:patterns}
\end{table*}

\section{Success Probability and Resource Requirements}
\label{sec:prob}

The analysis presented here focuses on the high success probability
and high success probability (extended) regimes noted in
Tab.~\ref{tab:link-epp}.  Let us now make the boundaries between the
probability regimes more concrete.  In order to achieve deterministic,
minimal timing over a path of $h$ hops, we want a high probability of
all of the hops succeeding in building enough entanglement.  Let the
probability $P_p$ be the probability that the entire path
\emph{cascades} successfully end to end on a given trial, and $P_s$ be
the probability that a single hop successfully cascades.  Obviously,
$P_p = (P_s)^h$ for homogeneous hops, or
\begin{align}
P_p = \prod_{j=1}^{h} P_s(j)
\end{align}
for heterogeneous hops; here, we will assume homogeneous. If we assume
$h = 50$ (long enough to cover 1,000km at 20km/hop), for $P_s = 0.99$,
we find $P_p \sim 0.6$.

\subsection{Asynchronous/Cascade Probability Boundary}

How many transmitters $N_T$ do we need in each node to achieve $P_s =
0.99$?  We need at least $n$ successful entanglements in one burst,
where $n$ is the block size we need for error correction (or $d$ for
the surface code).  The mean number of successful entanglements is
$N_r = P_rN_T$.  To guarantee that
\begin{align}
  P_s = P(N_r \ge n) = 1 - \sum_{j=0}^{n} {N_T \choose
    j}P_r^j(1-P_r)^{N_T-j} \ge 0.99,
\end{align}
for $P_r = 0.01$ and $n = 7$ (as in the Steane [[7,1,3]]
code~\cite{steane96:_qec}), we must have $N_T \ge 1460$, a rather
substantial overhead.  The ratio is similar for $0.001 < P_r < 0.1$
for the 7-qubit code, and somewhat better for the [[23,1,7]] Golay
code~\cite{PhysRevA.79.032325}.  We can say roughly, then, that the
lower bound for the ``high entanglement probability'' regime is $P_r
\sim 2n/N_T$.  Above this level, we can coordinate the activities on
links to create an efficient path, as discussed in this article.

\subsection{Cascade Generic/Direct Transmission Probability
  Boundary}
\label{sec:direct-prob}

Likewise, we can calculate the upper bound for ``high probability'',
the transition to \emph{direct transmission} of quantum data, without
first building generic states such as Bell states.  Although certain
cases allow reconstruction with less than half of the physical qubits
comprising a logically encoded state~\cite{silva2004erasure}, in
general, the no-cloning theorem shows that we need to receive more
than half, that is,
\begin{align}
  P_s = P(N_r \ge \lceil h/2\rceil) = 1 - \sum_{j=0}^{\lfloor
    h/2\rfloor} {n \choose j}P_r^j(1-P_r)^{n-j}.
\end{align}
For direct transmission we need $P_s$ to be substantially higher than
0.99, e.g., $P_s = 0.999$ or better when transporting valuable quantum
data, perhaps as high as $1-10^{-15}$ for some forms of distributed
quantum
computation~\cite{Chien:2015:FOU:2810396.2700248,van-meter14:_quantum_networking}.
For the seven-qubit Steane code and $P_s = 0.999$, we would need $P_r
\ge 0.93$, whereas for the 23-qubit Golay code we would need $P_r \ge
0.79$.  The value of $\alpha$ in Tab.~\ref{tab:link-epp} therefore is
highly code dependent, but for our purposes here can be considered to
be in the range $0.8 \le \alpha \le 0.95$.  We call the region $0.5 <
P_r \le \alpha$ the ``high probability (extended)'' regime, and in
this regime our timing analysis still applies.

Direct transmission requires that the polarity of the links correspond
to the desired direction of propagation along the entire path, or
equivalently that the links are duplex.  When this condition does not
hold, operational procedures will instead fall back to the procedures
and timing behavior discussed in this paper.


\section{Discussion}

Muralidharan \emph{et al.} divided quantum repeater architectures into
three generations~\cite{muralidharan15:_ultrafast-generations}:
\begin{enumerate}
\item First generation: using bidirectional entanglement purification
  (2-EPP) and acknowledged (heralded) entanglement generation (known
  as AEC or HEG);
\item Second generation: using unidirectional entanglement
  purification (1-EPP), e.g. based on error correcting codes, and
  acknowledged (sometimes called heralded) entanglement generation
  (known as AEC or HEG); and
\item Third generation: using  unidirectional entanglement
  purification (1-EPP), e.g. based on error correcting codes, and
  unacknowledged but heralded (without acknowledgement to partner)
  loss error management, allowing direct transmission of states.
\end{enumerate}

Our analysis differs in several respects: (1) we discuss the three
application classes B, C, and T, whereas earlier analysis focused on
the QKD application or equivalent Bell inequality violation tests; (2)
we focus on optimizing the timing pattern; (3) we explicitly consider
heterogeneity in paths; and (4) we study the impact of the link
polarity.  Primarily, our contribution can be considered to be
optimization of 2G repeater timing, but the B/C/T distinction allows
us to also further subdivide 2G repeaters into path-optimizable and
non-path-optimizable subclasses.

Even for the relatively simple useage of the Steane code, and assuming
improvement to around 1\% entanglement success probability, the
resource requirements for achieving this cascaded operation are
substantial.  1,400 transmitter qubits is likely to remain well out of
reach in actual implementation for a number of years.  Moreover, with
1,400 transmitters, the expected number of successful entanglements
would be $\sim 14$, enough to create two Steane transfers.  The more
likely operational approach, therefore, would be to continue
asynchronous operation in order to take advantage of the extra
resources.

Once entanglement success probability reaches several percent, the
optimizations presented here will guide network protocols and
operations to substantially improved performance by reducing buffer
memory needs (slightly at low $P_r$, substantially at high $P_r$), and
by reducing memory decoherence.  Following the guidelines here, the
location of buffer memory consumed can be adjusted along a Pareto
optimal frontier, providing operational flexiblity.  With end-to-end
latencies of around $T_E \approx 100$msec between some locations on
the globe, the reduction from $5T_E$ (worst case for ridge fold timing
on butterfly links) to $2T_E$ in the best case can save hundreds of
milliseconds of memory decoherence time in a Quantum Internet.

\section*{Acknowledgements}

\myindent The authors acknowledge useful discussions with Chip
Elliott, Simon Devitt and Joe Touch.  This work was supported by
JSPS Kakenhi Kiban B 16H02812.

\bibliographystyle{unsrt}
\bibliography{paper-reviews.bib} 

\end{document}